\documentclass[conference]{IEEEtran}

\IEEEoverridecommandlockouts

\usepackage{bm,cite,algorithm,algorithmic,float,amsmath,amssymb}
\usepackage[dvips]{graphicx}
\usepackage{subfigure,amsfonts}
\usepackage{cite, color}
\usepackage{balance}
\floatname{algorithm}{Algorithm}
\ifCLASSINFOpdf
\else
\fi
\begin{document}
%
\title{Power Splitting for Full-Duplex Relay\\ with Wireless Information and Power Transfer}

\author{\IEEEauthorblockN{Hongwu Liu}
\IEEEauthorblockA{Shandong Jiaotong University\\
Jinan, China\\
Email: hong.w.liu@hotmail.com}
\and
\IEEEauthorblockN{Kyeong Jin Kim}
\IEEEauthorblockA{Mitsubishi Electric Research Laboratories (MERL)\\
Cambridge, MA, USA\\
Email: kyeong.j.kim@hotmail.com}
\and
\IEEEauthorblockN{Kyung Sup Kwak}
\IEEEauthorblockA{TelLab, Inha University\\
Incheon, Korea\\
Email: kskwak@inha.ac.kr}}

\maketitle

\begin{abstract}
This paper investigates power splitting for full-duplex relay networks with wireless information and energy transfer. By applying power splitting as a relay transceiver architecture, the full duplex information relaying can be powered by energy harvested from the source-emitted radio frequency signal. In order to minimize outage probability, power splitting ratios have been dynamically optimized according to full channel state information (CSI) and partial CSI, respectively. Under strong loop interference, the proposed full CSI-based and partial CSI-based power splitting schemes achieve the better outage performance than the fixed power splitting scheme, whereas the partial CSI-based power splitting scheme can ensure competitive outage performance without requiring CSI of the second-hop link. It is also observed that the worst outage performance is achieved when the relay is located midway between the source and destination, whereas the outage performance of partial CSI-based power splitting scheme approaches that of full CSI-based scheme when the relay is placed close to the destination.
\end{abstract}

\IEEEpeerreviewmaketitle

\section{Introduction}

Energy harvesting (EH) has emerged as a promising enabling technology for wireless cooperative or sensor networks to function in environment with physical or economic limitations \cite{Energy_harvesting,Energy_harvesting_relay, SWIPT_protocol_AF}. Through EH from ambient radio-frequency (RF) signals, periodic battery replacement or recharging can be alleviated for energy-constrained sensor or relay nodes. Since RF signals can carry both information and energy, simultaneous wireless information and power transfer (SWIPT) has been proposed \cite{Varshney_SWIET,Grover_Shannon_meets_Tesla, SWIPT_cellular,SWIPT_magazine,SWIPT_protocol_DF,SWIPT_protocol_AF} and two practical receiver architectures, namely, time switching (TS) and power splitting (PS) \cite{SWIPT_architecture}, have been widely adopted in various SWIPT systems \cite{SWIPT_cellular, SWIPT_OFDM, MIMO_B_SWIPT}.

By employing TS-based relaying (TSR) and PS-based relaying (PSR) protocols for amplify-and-forward (AF) systems \cite{SWIPT_protocol_AF},  SWIPT can not only keep energy-constrained relay nodes active through RF EH, but also enable information relaying across barriers or over long distance. The outage and diversity performances of SWIPT for cooperative networks with spatially random relays were investigated in \cite{SWIPT_SRR} and the distributed PS-based SWIPT was designed for interference relay systems \cite{SWIPT_DPS}. Several power allocation schemes for EH relay systems with multiple source-destination pairs were investigated in \cite{SWIPT_PA}. Furthermore, antenna switching and antenna selection have also been applied for SWIPT relaying systems \cite{SWIPT_antenna_switch, SWIPT_Antenna_select}. Dynamic power splitting with full channel state information (CSI) and partial CSI has been investigated for AF half-duplex relaying networks \cite{SWIPT_AF_DPS2}.
Since full-duplex relay (FDR) can improve spectral efficiency significantly over half-duplex relay, wireless information and power transfer for FDR networks has drawn much attention \cite{SWIPT_FD_selfenergy,SWIPT_FD}. Through concurrent information relaying and EH via separated relay transmit and receive antennas, the authors of \cite{SWIPT_FD_selfenergy} proposed a self-interference immunizing FDR scheme. In \cite{SWIPT_FD}, the throughput of TSR protocol has been analyzed for FDR SWIPT systems, in which TS factor has been optimized for EH relay to maximize system throughput. Since PSR protocols outperform TSR protocols in various scenarios \cite{SWIPT_protocol_AF, SWIPT_protocol_DF, SWIPT_DPS}, we just focus our attention on PSR protocol for FDR networks. In this paper, we consider a wireless FDR network using PSR protocol to realize SWIPT. In order to minimize outage probability, power splitting ratios are optimized with full CSI and partial CSI, respectively.

\begin{figure}[htbp]
\begin{center}
\includegraphics[width=3.2in]{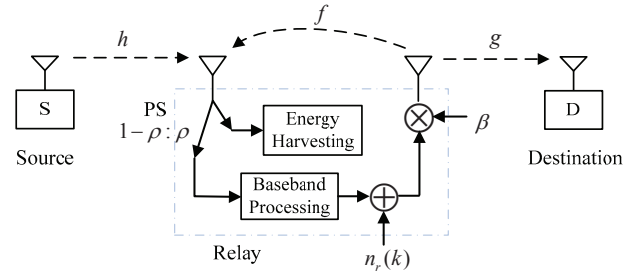}
\vspace{-0.09in}
\caption{Block diagram of the FDR network.}
\end{center}
\vspace{-0.15in}
\end{figure}

\section{System Model}

In the considered wireless FDR network, a source intends to transmit its information to a destination. Due to large separation between the source and destination, a AF relay operating in FDR mode is employed to relay the source information to the destination, as depicted in Fig. 1. For simplicity of implementation, the source and destination are equipped with single antenna, whreas the AF relay is equipped with single receive antenna and single transmit antenna. We assume that the initial relay running is supported by the initial energy stored in the battery. Then, the relay needs to harvest energy from its received RF signals to forward information. The channel from the source to relay and from the relay to destination are denoted by $h$ and $g$, respectively, whereas the loopback interference channel at the relay is denoted by $f$. We assume all the channels experience Raleigh fading and keep constant during each transmission block. The means of the exponential random variables $|h|^2$, $|g|^2$, and $|f|^2$ are denoted by $\lambda _h$, $\lambda _g$, and $\lambda _f$, respectively. By employing request-to-send/clear-to-send (RTS/CTS) based channel estimation scheme, the CSI can be estimated to facilitate the SWIPT \cite{SWIPT_architecture, SWIPT_FD_selfenergy,SWIPT_protocol_AF}. In this paper, we assume that the relay with the capability to access full CSI (or partial CSI) computes and updates the relay control parameters.

In each transmission block, the power of the relay-received signal is splitted in $\rho : 1- \rho$ proportion, where $\rho$ is the power splitting ratio. Since the relay operates in FDR mode, it concurrently receives the signal $y_r(t)$ and transmits the signal $x_r(t)$ on the same frequency.
The splitted signal at the input of the EH receiver is
\begin{eqnarray}
\sqrt{\tfrac{\rho P_s}{d_1^m}} hs(t) + \sqrt{\rho} f x_r(t),
\end{eqnarray}
where $s(t)$ is the source signal, $P_s$ is the source transmission power, $d_1$ is the distance between the source and relay, $m$ is the path loss exponent. The harvested energy at the relay can be expressed as
\begin{eqnarray}
{E_h} =\eta \rho \left( \tfrac{ {P_s}|h|^2}{{d_1^m}}  + |f|^2 P_r \right)T,
\end{eqnarray}
where $P_r = \mathbb{E}\{ |{x_r}(t){|^2}\}$ is the relay transmission power, $T$ is the duration time of each transmission block, and $\eta$ ($0< \eta <1$) is the energy conversion efficiency, which depends on the rectification process and the energy harvesting circuitry [6].
At the relay, the sampled baseband signal can be written as
\begin{eqnarray}
{y_r}(k) = \sqrt {\tfrac{{(1 - \rho ){P_s}}}{{d_1^m}}} hs(k) + \sqrt {1 - \rho } f{x_r}(k) + {n_r}(k), \label{eq:y_r}
\end{eqnarray}
where $k$ denotes the symbol index, $s(k)$ is the sampled $s(t)$, $x_r(k)$ is the sampled signal of $x_r(t)$, ${n_r}(k)$ is the zero mean additive white Gaussian noises (AWGN) with variance $\sigma ^2$.
In \eqref{eq:y_r}, the second term on the right hand side is the residual loop interference at the relay node. Using the harvested energy, the relay amplifies the received signal by a relay gain $\beta$. Then, the transmitted signal at the relay can be expressed as
\begin{eqnarray}
{x_r}(k) = \sqrt \beta  {y_r}(k - \tau ), \label{eq:x_r}
\end{eqnarray}
where $\tau \ge 1$ is the processing delay at the relay.
By recursively substituting \eqref{eq:y_r} and \eqref{eq:x_r}, we have the following expression for the transmitted signal at the relay:
\begin{eqnarray}
{x_r}(k) &=& \sqrt \beta  \sum\limits_{j = 1}^\infty  {{(\sqrt {(1 - \rho )\beta } f)}^{j - 1}} \nonumber \\
& & \times \left( {\sqrt {{\textstyle{{(1 - \rho ){P_s}} \over {d_1^m}}}} hs(k - j\tau ) + {n_r}(k - j\tau )} \right).  \label{eq:x_r_recursive}
\end{eqnarray}
The sampled received signal at the destination, $y_d(k)$ is given by
\begin{eqnarray}
{y_d}(k) = {\textstyle{1 \over {\sqrt {d_2^m} }}}g{x_r}(k) + {n_d}(k), \label{eq:y_d}
\end{eqnarray}
where $d_2$ is the distance from the relay to destination, ${n_d}(k)$ is the zero mean AWGN with variance $\sigma ^2$. Substituting \eqref{eq:x_r_recursive} into \eqref{eq:y_d}, we have
\begin{eqnarray}
\!\!\!\!{y_d}(k) &\!\!\!\!\!=\!\!\!\!\!& \sqrt {{\textstyle{{(1 - \rho ){P_s}\beta } \over {d_1^md_2^m}}}} gh\sum\limits_{j = 1}^\infty  {{{(\sqrt {(1 - \rho )\beta } f)}^{j - 1}}s(k - j\tau )}  \nonumber \\
\!\!\!\!&\!\!\!\!\!\!\!\!\!\!&  + \sqrt {{\textstyle{\beta  \over {d_2^m}}}} g\sum\limits_{j = 1}^\infty \!\! {{{(\sqrt {(1 - \rho )\beta } f)}^{j - 1}}\!\!{n_r}(k - j\tau )}  \!+\! {n_d}(k). \label{eq:y_d_recursive}
\end{eqnarray}

\subsection{End-to-End {\rm{SINR}}}

In the following, we derive the end-to-end signal power under the condition of employing non-oscillatory relay. By assuming that all the signal and noise samples are mutually independent, we calculate the relay transmission power $P_r  = \mathbb{E}\{ |{x_r}(k){|^2}\}$ from \eqref{eq:x_r_recursive} as
\begin{eqnarray}
P_r  &\!\!\!=\!\!\!& \beta \sum\limits_{j = 1}^\infty  {{((1 - \rho )\beta |f|^2 )}^{j - 1}}  \left( {{{(d_1^m)}^{ - 1}}(1 - \rho ){P_s}|h{|^2} + \sigma ^2} \right) \nonumber \\
&\!\!\! =\!\!\! & \beta \frac{{{{(d_1^m)}^{ - 1}}(1 - \rho ){P_s}|h{|^2} + \sigma ^2}}{{1 - (1 - \rho )|f{|^2}\beta }}.  \label{eq:P_r}
\end{eqnarray}
To prevent oscillation and guarantee finite relay transmission power, the relay gain is limited by
\begin{eqnarray}
{\beta} < \frac{1}{(1-\rho){|f{|^2}}}  \label{eq:beta2<1/f2}
\end{eqnarray}
Given the relay-harvested energy, the maximum relay transmission power is expressed as
\begin{eqnarray}
P_{\rm max} = \frac{{{E_h}}}{{T}} = \eta \rho \left( \tfrac{ {P_s}|h|^2}{{d_1^m}}  + |f|^2 P_r \right). \label{eq:P_max}
\end{eqnarray}
The actual relay transmission power should be less than or equal to the maximum relay transmit power, i.e.,
\begin{eqnarray}
P_{r} \le P_{\rm max}. \label{eq:P_r<P_max}
\end{eqnarray}
When \eqref{eq:P_r} and \eqref{eq:P_max} are substituted into \eqref{eq:P_r<P_max}, the relay gain under the maximum relay transmission power is limited by
\begin{eqnarray}
\beta \le \frac{{\eta \rho {\gamma _{{\rm{SR}}}}}}{{(1 - \rho ){\gamma _{{\rm{SR}}}} - \eta \rho |f{|^2} + 1}}, \label{eq:beta2<beta_max}
\end{eqnarray}
where the signal-to-noise ratio (SNR) of the source-relay channel is defined as $\gamma _{\rm SR} \triangleq {{P_s |h|^2}\over {d_1^m \sigma ^2}}$. Moreover, the non-oscillatory condition \eqref{eq:beta2<1/f2} is also guaranteed by \eqref{eq:beta2<beta_max}.
At symbol index $k$, the destination node can employ any standard detection procedure to decode the desired signal $s(k-\tau)$, and the rest of the received signal components act as interference and noise. Based on the assumption that signal and noise are independent of each other, the received signal power at the destination node is calculated from \eqref{eq:y_d} as ${\mathbb{E}} \{|y_d(k)|^2 \}= (d_2^m)^{-1} |g|^2 {\mathbb{E}}\{|x_r(k)|^2\}+ \sigma ^2$,
which can be further evaluated as follows, comprising of the desired signal power, loop interference power, and noise power:
\begin{eqnarray}
& & \!\!\!\!\!\!\!\!\!\!\!\! {\mathbb{E}} \{ |{y_d}(k){|^2}\}  = \underbrace {{{(d_1^md_2^m)}^{ - 1}}(1 - \rho )\beta  {P_s}|h{|^2} |g{|^2}}_{{\rm{desired}}\;{\rm{signal}}\;{\rm{power}}} \nonumber \\
& &\!\!\!\!\!\! + \underbrace {\left( {{{(d_1^m)}^{ - 1}}(1 - \rho ){P_s}|h{|^2} + {\sigma ^2}} \right)\beta {{(d_2^m)}^{ - 1}}|g{|^2}\tfrac{{(1 - \rho )\beta |f{|^2}}}{{1 - (1 - \rho )\beta |f{|^2}}}}_{{\rm{loop}}\;{\rm{interference}}\;{\rm{power}}}   \nonumber \\
& &\!\!\!\! + \underbrace {(\beta {{(d_2^m)}^{ - 1}}|g{|^2} + 1){\sigma ^2}}_{{\rm{noise}}\;{\rm{power}}}   \label{eq:y_d_power}
\end{eqnarray}
Based on \eqref{eq:y_d_power}, the end-to-end signal-to-interference-plus-noise-ratio (e-SINR) at the destination is given by
\begin{eqnarray}
\gamma  = \frac{{(1 - \rho ){\gamma _{{\rm{SR}}}}{\gamma _{{\rm{RD}}}}}}{{{\gamma _{{\rm{SR}}}}/\beta  + {\gamma _{{\rm{RD}}}} + ((1 - \rho ){\gamma _{{\rm{SR}}}} + 1){\gamma _{{\rm{RD}}}}\frac{{(1 - \rho )|f{|^2}}}{{1/\beta  - (1 - \rho )|f{|^2}}}}},  \label{eq:SINR}
\end{eqnarray}
where the SNR of the relay-destination channel is defined as ${\gamma _{{\rm{RD}}}} \triangleq \tfrac{{{P_s}|h{|^2}|g{|^2}}}{{d_1^md_2^m{\sigma ^2}}}$.

\section{Power Splitting with Full CSI}

In this section, we assume that full CSI is available at the relay and investigate how to compute the relay gain and power splitting ratio.

According to \eqref{eq:y_d_power} and \eqref{eq:SINR}, the e-SINR has a very complicated non-linear relationship with $\beta$ and $\rho$.
The design goal of the relay control with full CSI is to minimize outage probability by optimizing the control parameters $\{\beta, \rho\}$, which in turn is to maximize e-SINR. Therefor, the optimal $\{ \beta ^*, \rho ^*\}$ can be obtained by solving the following optimization problem
\begin{eqnarray}
& &~~~~~~~~~~\{ {\beta ^*},{\rho ^*}\}  = \arg \mathop {\max }\limits_{\beta ,\rho } \gamma  \label{eq:P1}\\
& &{\rm s.t.}~~~ 0< \beta  \le \tfrac{{\eta \rho {\gamma _{{\rm{SR}}}}}}{{(1 - \rho ){\gamma _{{\rm{SR}}}} - \eta \rho |f{|^2} + 1}} ~~{\rm and}~~ 0<\rho<1.  \nonumber
\end{eqnarray}
Since $\gamma $ is not jointly concave in $\beta$ and $\rho$, the optimal $\{ \beta ^* ,\rho ^*\}$  in \eqref{eq:P1} can be obtained by exhaustive searching for all the possible numerical combinations of $\{\beta ,\rho \}$.

In this work, we adopt a simple and popular relay gain by setting the relay gain at the maximum relay transmission power. For a given $\rho$ with any value in the range $(0, 1)$, the relay-harvested energy and maximum relay transmission power are determined,
so that the relay gain is given by
\begin{eqnarray}
{\beta  } = \frac{{\eta \rho {\gamma _{{\rm{SR}}}}}}{{(1 - \rho ){\gamma _{{\rm{SR}}}} - \eta \rho |f{|^2} + 1}}.  \label{eq:beta_max}
\end{eqnarray}
By substituting \eqref{eq:beta_max} into \eqref{eq:SINR}, the e-SINR can be expressed as
\begin{eqnarray}
{\gamma  } = \frac{{\varepsilon \eta \rho (1 - \rho ){\gamma _{{\rm{SR}}}}{\gamma _{{\rm{RD}}}}}}{{(1 - \rho )(\varepsilon  + {\eta ^2}{\rho ^2}{\gamma _{{\rm{RD}}}}|f{|^2}){\gamma _{{\rm{SR}}}} + \varepsilon (\varepsilon  + \eta \rho {\gamma _{{\rm{RD}}}})}}, \label{eq:SINR_max}
\end{eqnarray}
where $\varepsilon \triangleq 1 - \eta \rho |f{|^2}$.
Now, the maximization of $\gamma  $ becomes an optimization problem of $\rho$.
Since $\gamma $ is concave with respect to $\rho$, the optimized power splitting ratio that maximizes $\gamma  $ can be obtained by solving $\frac{{\partial {\gamma  }}}{{\partial \rho }} = 0$. Furthermore, $\frac{{\partial {\gamma  }}}{{\partial \rho }} = 0$ can be simplified as $Q_1(\rho) =0$, where
\begin{eqnarray}
Q_1(\rho) \triangleq a_4 \rho ^4+a_3 \rho ^3+a_2 \rho ^2+a_1 \rho +a_0
\end{eqnarray}
is a quartic equation and
\begin{subequations}
\begin{eqnarray}
\!\! \!\!\!\! \!\! \!\!  a_0 &\!\!\!\!\!=\!\!\!\!\!& 1  +  \gamma _{\rm SR}, \\
\!\! \!\!\!\! \!\! \!\! a_1 &\!\!\!\!\!=\!\!\!\!\!& -2(1 + \eta |f|^2)(1  +  \gamma _{\rm SR}),\\
\!\!\!\!\!\! \!\! \!\!   a_2 &\!\!\!\!\!=\!\!\!\!\!& {\gamma _{{\rm{SR}}}} - \eta {\gamma _{{\rm{RD}}}} + {\eta ^2}|f{|^4}(1 + {\gamma _{{\rm{SR}}}}) \nonumber \\
\!\!\!\! \!\! \!\! \!\!  &\!\!\!\!\! \!\!\!\!\!& ~~~~~~~~~~~~~~~+ \eta |f{|^2}(5 + {\gamma _{{\rm{SR}}}}(4 - \eta {\gamma _{{\rm{RD}}}})), \\
\!\!\!\!\!\! \!\! \!\!   a_3 &\!\!\!\!\!=\!\!\!\!\!&  - 2 \eta |f|^2 (\gamma _{\rm SR} + \eta |f|^2 (2 + \gamma _{\rm SR}) \!-\! \eta (1 + \gamma _{\rm SR})\gamma _{\rm RD}),\\
\!\!\!\!\!\! \!\! \!\!   a_4 &\!\!\!\!\!=\!\!\!\!\!& {\eta ^2} |f|^2 (\eta |f|^2  + \gamma _{\rm SR})(|f|^2 - \gamma _{\rm RD}) .
\end{eqnarray}
\end{subequations}
After some mathematical manipulations, we can calculate the optimized power splitting ratio by
\begin{eqnarray}
\rho  \!=\! \left\{\!\!\! {\begin{array}{*{20}{c}}
{{\rm{the~1st~root~of~}}Q_1(\rho ) = 0,}&{\!\!{|f|^2 \ge \gamma _{{\rm{RD}}}}}\\
{{\rm{the~2nd~root~of~}}Q_1(\rho ) = 0,}&{\!\!|f{|^2} <  \gamma _{\rm{RD}}}
\end{array}} \right. .\!\!\!\!\!  \label{eq:rho_max}
\end{eqnarray}
The solution formula of a quartic equation can be obtained by using Descartes-Euler-Cardano¡¯s method or Ferrari-Lagrange¡¯s \cite{Quartic_Equations}.

\section{Power Splitting with Partial CSI}

Since the CSI of the second-hop link can be estimated only at the destination side, the signaling exchanging for estimating $g$ becomes a heavy burden for the power splitting scheme with full CSI. In this section, we consider the power splitting ratio optimization with partial CSI, i.e., the CSI of the first-hop and loop interference channels is available at the relay.

The optimization problem of the power splitting ratio is to find a optimal $\rho ^*$ to minimize the outage probability conditioned on partial CSI, which can be written as
\begin{eqnarray}
\rho ^* = \arg \mathop {\min }\limits_{0< \rho <1 }  \Pr (\gamma (\rho) < \gamma _0 | h, f ), \label{eq:opt_problem}
\end{eqnarray}
where $\gamma _0$ is the target e-SINR ($\gamma _0 > 0$).

Substituting \eqref{eq:SINR_max} into $P _{\rm out} = \Pr (\gamma   < \gamma _0 | h, f)$, the conditioned outage probability  is given by
\begin{eqnarray}
{P_{{\rm{out}}}} = \Pr \left( {\left. {|g{|^2} < \tfrac{{{G_1}(\rho )}}{{{G_2}(\rho )}}} \right|h,f} \right), \label{eq:p_out_max}
\end{eqnarray}
where ${G_1}(\rho ) = a|h{|^2} + b$, ${G_2}(\rho ) = c|h{|^4} + d|h{|^2}$, and
\begin{subequations}
\begin{eqnarray}
a &=& {P_s}d_1^md_2^m{\sigma ^2}{\gamma _0}(1 - \rho )(1 - \eta \rho |f{|^2}), \\
b &=& d_1^{2m}d_2^m\sigma ^4{\gamma _0}{(1 - \eta \rho |f{|^2})^2}, \\
c &=& \eta \rho P_s^2(1 - \rho )(1 - \eta \rho (1 + {\gamma _0})|f{|^2}), \\
d &=& {P_s}d_1^m{\sigma ^2}\eta \rho {\gamma _0}(\eta \rho |f{|^2} - 1).
\end{eqnarray}
\end{subequations}
Since $|g|^2$ is always greater than a negative number, $P_{\rm out}$ in \eqref{eq:p_out_max} becomes 1 when $\frac{{{G_1}(\rho )}}{{{G_2}(\rho )}}$ achieves a negative value. Therefore, by observing the sign of $\frac{{{G_1}(\rho )}}{{{G_2}(\rho )}}$ under the constraints of $0<\rho<1$ and $0<\eta < 1$, $P_{\rm out}$ can be simplified as
\begin{eqnarray}
{P_{{\rm{out}}}} \!\!=\!\! \left\{ \!\!\!\! {\begin{array}{*{20}{c}}
{\Pr \left( {\left. {|g{|^2} < \frac{{{G_1}(\rho )}}{{{G_2}(\rho )}}} \right|h,f} \right),}\!\!\!\!&\!\!\!{|f{|^2} < {F_1}{\rm{~and~}}|h{|^2} > {H_1}}\\
{\Pr \left( {\left. {|g{|^2} < \frac{{{G_1}(\rho )}}{{{G_2}(\rho )}}} \right|h,f} \right),}\!\!\!\!&\!\!\! \begin{array}{l}
~~~~|f{|^2} > {F_2}{\rm{~and~}}\\
(|h{|^2} < {H_1}{\rm{~or~}}|h{|^2} > {H_2})
\end{array}\\
\!\!\!\!\!\!\!\!\!\!\!\!\!\!\!\!1,\!\!\!\!\!\!&\!\!\!\!{|f{|^2} < {F_1}{\rm{~and~}}|h{|^2} < {H_1}}\\
\!\!\!\!\!\!\!\!\!\!\!\!\!\!\!\!1,\!\!\!\!\!\!&\!\!\!\!\!\!\!\!\!\!\!{{F_1} < |f{|^2} < {F_2}{\rm{~and~}}|h{|^2} > 0}\\
\!\!\!\!\!\!\!\!\!\!\!\!\!\!\!\!1,\!\!\!\!\!\!&\!\!\!\!\!\!\!\!\! {|f{|^2} > {F_2}{\rm{~and~}}{H_1} < |h{|^2} < {H_2}}
\end{array}} \right.\!\!\!\!\!\!, \nonumber
\end{eqnarray}
\vspace{-0.22in}
\begin{eqnarray}
\label{eq:p_out_max_2}
\end{eqnarray}
where ${H_1} \triangleq \frac{{d_1^m{\sigma ^2}{\gamma _0}(1 - \eta \rho |f{|^2})}}{{{P_s}(1 - \rho )(1 - \eta \rho |f{|^2}(1 + {\gamma _0}))}}$,  ${H_2} \triangleq \frac{{d_1^m{\sigma ^2}(1 - \eta \rho |f{|^2})}}{{{P_s}(\rho  - 1)}}$, ${F_1} \triangleq \frac{1}{{\eta \rho (1 + {\gamma _0})}}$, and ${F_2} \triangleq \frac{1}{{\eta \rho }}$.

In \eqref{eq:p_out_max_2}, when $P_{\rm out} = 1$, we can set $\rho =1$ such that the relay can harvest energy as much as possible. When  $P_{\rm out} = \Pr \left( {\left. {|g{|^2} < \tfrac{{{G_1}(\rho )}}{{{G_2}(\rho )}}} \right|h,f} \right)$, the design goal is to find a power splitting ratio to minimize the conditioned outage probability.
By substituting $0< \rho <1$ in $F_1$ and $H_1$, the CSI constraint $\{|f{|^2} < {F_1}{\rm{~and~}}|h{|^2} > {H_1}\}$ can be rewritten as
\begin{eqnarray}
\rho < \rho _1 {\rm{~and~}} C_1,
\end{eqnarray}
where ${\rho _1} =  - \frac{{\sqrt {{{(\eta {\gamma _0}|f{|^2} - {\gamma _{{\rm{SR}}}}(\eta |f{|^2}(1 + {\gamma _0}) - 1))}^2} + 4\eta {\gamma _{{\rm{SR}}}}\gamma _0^2|f{|^2}} }}{{2\eta {\gamma _{{\rm{SR}}}}|f{|^2}(1 + {\gamma _0})}}  + \frac{{{\gamma _{{\rm{SR}}}} - \eta {\gamma _0}|f{|^2}}}{{2\eta {\gamma _{{\rm{SR}}}}|f{|^2}(1 + {\gamma _0})}} + \frac{1}{2}$ and $C_1$ represents for $\{|f|^2< \tfrac{1}{\eta (1+\gamma _0)} {\rm{~and~}} \gamma _{\rm SR} > \gamma _0 \} $. Thus, when the CSI satisfies the constraint $C_1$, we should find the optimized power splitting ratio in the set $\Omega_1 = \{ \rho | 0<\rho <\rho _1 \}$. Similarly, when the CSI satisfies the constraints $C_2$: $\{ |f|^2 > \frac{1}{\eta}  {\rm{~and~}}   \gamma _{\rm SR} < \gamma _0 \}$ or $C_3$: $\{ |f|^2 > \frac{1}{\eta} {\rm{~and~}} \gamma _{\rm SR} > \eta |f|^2 -1 \}$, we should find the optimized power splitting ratio in the set $\Omega_2 = \{ \rho | \rho _1<\rho< 1 \} $ or $\Omega_3 = \{ \rho | \frac{1}{\eta |f|^2} <\rho< \frac{1+\gamma _{\rm SR}}{\gamma _{\rm SR} + \eta |f|^2} \} $.
Based on these observations, when the CSI satisfies the constraint $C_i$ ($i=1, 2$, and $3$), the conditioned outage probability can be computed by
\begin{eqnarray}
{\left. {\Pr \left( {\left. {\gamma _{\rm max} (\rho ) < {\gamma _0}} \right|h,f} \right)} \right|_{\rho  \in {\Omega _i}}} &\!\!\!\!=\!\!\!\!& \Pr \left( {\left. {|g{|^2} < \tfrac{{{G_1}(\rho )}}{{{G_2}(\rho )}}} \right|h,f} \right)  \nonumber
\end{eqnarray}
\begin{eqnarray}
\vspace{-0.2in}
\!\!\!\!\!\!\!\!\!\!\!\!\!\!\!\!\!\!\!\!\!\!\!\!\!\!\!\!\!\!\!\!&\!\!\!\!=\!\!\!\!&  \frac{1}{{{\lambda _g}}}\int_0^{\frac{{{G_1}(\rho )}}{{{G_2}(\rho )}}} {{e^{ - \frac{x}{{{\lambda _g}}}}}{\rm{d}}x} \nonumber \\
\!\!\!\!\!\!\!\!\!\!\!\!\!\!\!\!\!\!\!\!\!\!\!\!\!\!\!\!\!\!\!\! &\!\!\!\!=\!\!\!\!& 1 - \exp \left( { - \tfrac{{{G_1}(\rho )}}{{{G_2}(\rho ){\lambda _g}}}} \right) .
\end{eqnarray}
Since minimizing $1-\exp \left( { - \frac{{{G_1}(\rho )}}{{{G_2}(\rho ){\lambda _g}}}} \right)$ is equivalent to maximize $G (\rho) \triangleq \frac{{{G_2}(\rho )}}{{{G_1}(\rho )}}$, the optimized power splitting ratio can be obtained by solving
\begin{eqnarray}
\mathop {\rm Maximize }\limits_{\rho \in \Omega _i } ~~G (\rho).
\end{eqnarray}
As we can see that $G (\rho)$ is concave with respect to $\rho$ in the set $\Omega _i$ ($i=1, 2$, and 3), the solution $\rho ^*$ can be obtained by solving $\frac{{\partial G (\rho )}}{{\partial \rho }} = 0$. Unfortunately, given the complicated expression for $\frac{{\partial G (\rho )}}{{\partial \rho }} =0$, a closed-form solution is difficult to obtain. However, at the high SINRs, $G (\rho)$ has an approximation as $\tilde G (\rho) = \frac{{c|h{|^2} + d}}{{a}}$, so that an approximation of the optimized power splitting ratio can be obtained by solving $\frac{{\partial \tilde G (\rho )}}{{\partial \rho }} =0$. After some mathematical manipulations, the solution of $\frac{{\partial \tilde G (\rho )}}{{\partial \rho }} =0$ is given by
\begin{eqnarray}
\rho  \!=\! \left\{\!\!\! {\begin{array}{*{20}{c}}
{{\rm{the~1st~root~of~}}Q_2(\rho )=0,}&{\!\! \rho \in \Omega _1 {\rm ~and~} C_1}\\
{{\rm{the~2nd~root~of~}}Q_2(\rho )=0,}&{\!\! \rho \in \Omega _2 {\rm ~and~} C_2} \\
{{\rm{the~3rd~root~of~}}Q_2(\rho )=0,}&{\!\! \rho \in \Omega _3 {\rm ~and~} C_3}
\end{array}} \right. ,\!\!\!\!\!
\end{eqnarray}
where $Q_2(\rho) = c_4 \rho ^4 + c_3 \rho ^3 + c_2 \rho ^2 + c_1 \rho + c_0$ is a quartic function and
\begin{subequations}
\begin{eqnarray}
\!\!\!\!{c_0} &\!\!=\!\!& {\gamma _{{\rm{SR}}}} - {\gamma _0}, \\
\!\!\!\!{c_1} &\!\!=\!\!& 2\eta {\gamma _0}|f{|^2} - 2{\gamma _{{\rm{SR}}}}(1 + \eta |f{|^2}(1 + {\gamma _0})), \\
\!\!\!\!{c_2} &\!\!=\!\!& {\gamma _{{\rm{SR}}}}(1 + \eta |f{|^2}(4 + \eta |f{|^2}(1 + {\gamma _0})) - {\eta ^2}{\gamma _0}|f{|^4}\!\!,  \\
\!\!\!\!{c_3} &\!\!=\!\!&  - 2\eta {\gamma _{{\rm{SR}}}}|f{|^2}(1 + \eta |f{|^2})(1 + {\gamma _0}),  \\
\!\!\!\!{c_4} &\!\!=\!\!& {\eta ^2}{\gamma _{{\rm{SR}}}}|f{|^4}(1 + {\gamma _0}).
\end{eqnarray}
\end{subequations}
In computing the above expression of the power splitting ratio, the relay is required to know $h$ and $f$. By employing RTS/CTS based channel estimation scheme, the power splitting ratio can be determined before the transmission.

\vspace{-0.03in}
\section{Simulation Results}

This section presents some simulation results to verify the proposed power splitting schemes. In the simulation, the source transmission rate is $R = 3$ bps/Hz and the e-SINR threshold causing outage is given by $\gamma _0 = 2^R-1$. The energy harvesting efficiency is set to be $\eta =0.4$. The means of the channel gains are set as $\lambda _h = \lambda _g =1$, whereas the average interference-to-noise-ratio (INR) for the loopback channel is set as $\gamma _{\rm LI} \triangleq  \tfrac{\lambda _f}{\sigma ^2}$. The source transmission SNR is defined as ${\rm{SNR}} \triangleq P_s/\sigma ^2$. The path loos exponent is set to be $m=3$. Unless otherwise stated, the distance $d_1$ and $d_2$ are normalized to unit value. The fixed power splitting scheme proposed in \cite{SWIPT_protocol_AF} has also been simulated with $\rho =0.3$, $\rho =0.5$, and $\rho =0.7$ for comparison purposes.

\begin{figure}[htbp]
\begin{center}
\includegraphics[width=2.9in]{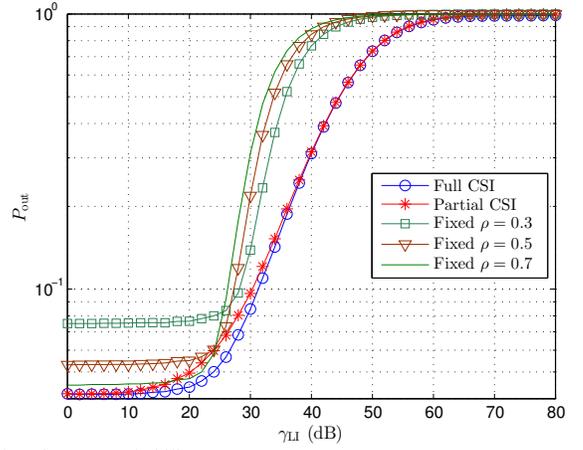}
\vspace{-0.15in}
\caption{Outage probability versus ${\gamma _{\rm LI}}$.}
\end{center}
\label{fig:2}
\end{figure}

\begin{figure}[htbp]
\begin{center}
\includegraphics[width=2.9in]{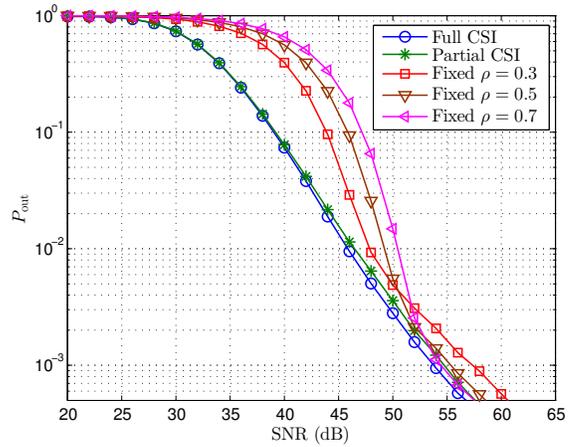}
\vspace{-0.15in}
\caption{Outage probability versus ${\rm{SNR}}$.}
\end{center}
\label{fig:3}
\end{figure}

Fig. 2 illustrates the impact of INR on the outage probability. In Fig. 2, the source transmission $\rm{SNR}$ is as 35 dB. In the practical FDR systems, the relay suffers from serious loop interference and the system performance degrades dramatically. As a result, the outage probability of the fixed power splitting scheme increases very quickly in the region of high INR. The proposed two power splitting schemes achieve the better performances than all the schemes with fixed $\rho$. For example, at the outage level of $10^{-1}$, the full CSI-based power splitting scheme achieves about 5.5 dB INR gain than the scheme with fixed $\rho =0.7$, about 4.5 dB INR gain than the scheme with fixed $\rho=0.5$, and about 3.5 dB INR gain than the scheme with fixed $\rho =0.3$. Although the full CSI-based power splitting scheme outperforms the partial CSI-based power splitting scheme in the region of middle and low INR, the performance gap is slight. This indicates that the partial CSI-based power splitting scheme, which incurs less overhead, approaches the full CSI-based scheme closely.

\begin{figure}[htbp]
\begin{center}
\includegraphics[width=2.9in]{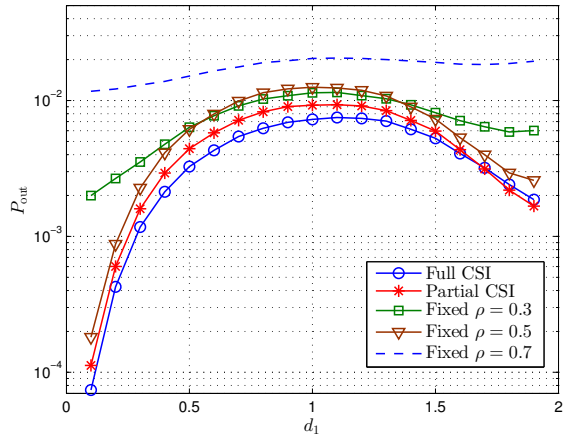}
\vspace{-0.15in}
\caption{Outage probability versus $d_1$.}
\end{center}
\label{fig:4}
\vspace{-0.15in}
\end{figure}

Fig. 3 shows the outage probability versus the source transmission $\rm{SNR}$. In the evaluation of Fig. 3, we set $\gamma _{\rm LI} =40$ dB. The proposed full CSI-based power splitting scheme achieves the best outage performance.
For example, at the outage probability level of $10^{-2}$, the full CSI-based power splitting scheme obtains about 4.5 dB gain than the scheme with fixed $\rho=0.7$, 3 dB gain than the scheme with fixed $\rho =0.5$, and 1.8 dB gain than the scheme with fixed $\rho =0. 3$. However, when the source transmission $\rm{SNR}$ becomes very high, the gain gap between the fixed power splitting scheme and proposed power splitting schemes decrease to a certain value.
Moreover, the partial CSI-based power splitting scheme achieves a outage performance very close to that of the full CSI-based scheme. Since the partial CSI-based scheme requires no knowledge of the second-hop link, it is more suitable for the system with less overhead.

Fig. 4 illustrates the outage probability versus the relay location. In the evaluation, we set $\rm{SNR} = 45$ dB, $\gamma _{\rm LI} = 35$ dB, $d_1+d_2=2$, and $d_1$ varies from 0.1 to 0.9. As shown in Fig. 4, for all the schemes, the highest outage probability is obtained. Thus, the relay should be placed near the source or destination to reduce the outage.
The proposed full CSI-based scheme achieves the best outage performance. However, Fig. 4 also shows that the partial CSI-based scheme approaches the outage performance of the full CSI-based scheme when the relay is placed near the destination. Since that we have applied a high SINR assumption in determining the power splitting ratio for the full CSI-based scheme, the partial CSI-based scheme even outperforms a little over the full CSI-based scheme in this case. The reason for this scenario is that the e-SINR is mainly determined by both the harvested energy and loop interference when the relay is placed far from the source, and the partial CSI-based power splitting scheme optimizes the power splitting ratio just based on the CSI of the first-hop and loopback channels. This result suggests that the partial CSI-based scheme is more preferable when the relay is placed near the destination.

\section{Conclusion}
In this study, full CSI-based and partial CSI-based power splitting schemes have been proposed for the FDR network with wireless information and power transfer. The power splitting ratio has been optimized to minimize outage probability under loop interference effect. Through simulations, it is found that both proposed power splitting schemes outperforms the fixed power splitting scheme. The partial CSI-based power splitting scheme achieves a competitive outage performance with less overhead over the full CSI-based power splitting scheme. It is also found that the worst outage performance is obtained by the relay placed midway between the source and destination, whereas the partial CSI-based power splitting scheme approaches the outage performance of the full CSI-based scheme when the relay is placed close to the destination.

\bibliography{IEEEabrv,IEEE_bib}

\begin{thebibliography}{10}
\providecommand{\url}[1]{#1}
\csname url@samestyle\endcsname
\providecommand{\newblock}{\relax}
\providecommand{\bibinfo}[2]{#2}
\providecommand{\BIBentrySTDinterwordspacing}{\spaceskip=0pt\relax}
\providecommand{\BIBentryALTinterwordstretchfactor}{4}
\providecommand{\BIBentryALTinterwordspacing}{\spaceskip=\fontdimen2\font plus
\BIBentryALTinterwordstretchfactor\fontdimen3\font minus
  \fontdimen4\font\relax}
\providecommand{\BIBforeignlanguage}[2]{{%
\expandafter\ifx\csname l@#1\endcsname\relax
\typeout{** WARNING: IEEEtran.bst: No hyphenation pattern has been}%
\typeout{** loaded for the language `#1'. Using the pattern for}%
\typeout{** the default language instead.}%
\else
\language=\csname l@#1\endcsname
\fi
#2}}
\providecommand{\BIBdecl}{\relax}
\BIBdecl

\bibitem{Energy_harvesting}
W.~K.~G. Seah, Z.~A. Eu, and H.~P. Tan, ``Wireless sensor networks powered by
  ambient energy harvesting ({WSN-HEAP}) - survey and challenges,'' in
  \emph{Proc. Wireless VITAE 2009}, Aalborg, Denmark, May 2009, pp. 1--5.

\bibitem{Energy_harvesting_relay}
C.~Huang, R.~Zhang, and S.~Cui, ``Throughput maximization for the gaussian
  relay channel with energy harvesting constraints,'' \emph{IEEE J. Sel. Areas
  in Commun.}, vol.~31, no.~8, pp. 1469--1479, Aug. 2013.

\bibitem{SWIPT_protocol_AF}
A.~Nasir, X.~Zhou, S.~Durrani, and R.~Kennedy, ``Relaying protocols for
  wireless energy harvesting and information processing,'' \emph{IEEE Trans.
  Wireless Commun.}, vol.~12, no.~7, pp. 3622--3636, Jul. 2013.

\bibitem{Varshney_SWIET}
L.~R. Varshney, ``Transporting information and energy simultaneously,'' in
  \emph{Proc. IEEE ISIT 2008}, Toronto, Canada, July 2008, pp. 1612--1616.

\bibitem{Grover_Shannon_meets_Tesla}
P.~Grover and A.~Sahai, ``Shannon meets tesla: Wireless information and power
  transfer,'' in \emph{Proc. IEEE ISIT 2010}, Austin, Tx, June 2010, pp.
  2363--2367.

\bibitem{SWIPT_cellular}
K.~Huang and V.~Lau, ``Enabling wireless power transfer in cellular networks:
  Architecture, modeling and deployment,'' \emph{IEEE Trans. Wireless Commun.},
  vol.~13, no.~2, pp. 902--912, Feb. 2014.

\bibitem{SWIPT_magazine}
I.~Krikidis, S.~Timotheou, S.~Nikolaou, G.~Zheng, D.~Ng, and R.~Schober,
  ``Simultaneous wireless information and power transfer in modern
  communication systems,'' \emph{IEEE Commun. Mag.}, vol.~52, no.~11, pp.
  104--110, Nov. 2014.

\bibitem{SWIPT_protocol_DF}
A.~Nasir, X.~Zhou, S.~Durrani, and R.~Kennedy, ``Throughput and ergodic
  capacity of wireless energy harvesting based df relaying network,'' in
  \emph{Proc. IEEE ICC 2014}, Sydney, Australia, June 2014, pp. 4066--4071.

\bibitem{SWIPT_architecture}
X.~Zhou, R.~Zhang, and C.~K. Ho, ``Wireless information and power transfer:
  Architecture design and rate-energy tradeoff,'' \emph{IEEE Trans. Commun.},
  vol.~61, no.~11, pp. 4754--4767, Nov. 2013.

\bibitem{SWIPT_OFDM}
D.~W.~K. Ng, E.~S. Lo, and R.~Schober, ``Energy-efficient resource allocation
  in multiuser ofdm systems with wireless information and power transfer,'' in
  \emph{Proc. IEEE WCNC 2013}, Shanghai, China, April 2013, pp. 3823--3828.

\bibitem{MIMO_B_SWIPT}
R.~Zhang and C.~K. Ho, ``{MIMO} broadcasting for simultaneous wireless
  information and power transfer,'' \emph{IEEE Trans. Wireless Commun.},
  vol.~12, no.~5, pp. 1989--2001, May 2013.

\bibitem{SWIPT_SRR}
Z.~Ding, I.~Krikidis, B.~Sharif, and H.~V. Poor, ``Wireless information and
  power transfer in cooperative networks with spatially random relays,''
  \emph{IEEE Trans. Wireless Commun.}, vol.~13, no.~8, pp. 4440--4453, Aug.
  2014.

\bibitem{SWIPT_DPS}
H.~Chen, Y.~Li, Y.~Jiang, Y.~Ma, and B.~Vucetic, ``Distributed power splitting
  for {SWIPT} in relay interference channels using game theory,'' \emph{IEEE
  Trans. Wireless Commun.}, vol.~14, no.~1, pp. 410--420, Aug. 2014.

\bibitem{SWIPT_PA}
Z.~Ding, S.~M. Perlaza, I.~Esnaola, and H.~V. Poor, ``Power allocation
  strategies in energy harvesting wireless cooperative networks,'' \emph{IEEE
  Trans. Wireless Commun.}, vol.~13, no.~2, pp. 846--860, Feb. 2014.

\bibitem{SWIPT_antenna_switch}
I.~Krikidis, S.~Sasaki, S.~Timotheou, and Z.~Ding, ``A low complexity antenna
  switching for joint wireless information and energy transfer in mimo relay
  channels,'' \emph{IEEE Trans. Commun.}, vol.~62, no.~5, pp. 1577--1587, May
  2014.

\bibitem{SWIPT_Antenna_select}
Z.~Zhou, M.~Peng, Z.~Zhao, and Y.~Li, ``Joint power splitting and antenna
  selection in energy harvesting relay channels,'' \emph{IEEE Trans. Signal
  Process.}, vol.~22, no.~7, pp. 823--827, Jul. 2015.

\bibitem{SWIPT_AF_DPS2}
L.~Hu, C.~Zhang, and Z.~Ding, ``Dynamic power splitting policies for {AF} relay
  networks with wireless energy harvesting,'' in \emph{Proc. IEEE ICC 2015},
  London, UK, 8-12, June 2015, pp. 1--5.

\bibitem{SWIPT_FD_selfenergy}
Y.~Zeng and R.~Zhang, ``Full-duplex wireless-powered relay with self-energy
  recycling,'' \emph{IEEE Wireless Commun. Lett.}, vol.~PP, no.~99, pp. 1--1,
  2015.

\bibitem{SWIPT_FD}
C.~Zhong, H.~Suraweera, G.~Zheng, I.~Krikidis, and Z.~Zhang, ``Wireless
  information and power transfer with full duplex relaying,'' \emph{IEEE Trans.
  Commun.}, vol.~62, no.~10, pp. 3447--3461, Oct. 2014.

\bibitem{Quartic_Equations}
S.~Neumark, \emph{Solution of Cubic and Quartic Equations}.\hskip 1em plus
  0.5em minus 0.4em\relax Oxford, NY: Pergamon Press, 1965.

\end{thebibliography}

\end{document}